# Wildfires and Climate Change: What does social media tell us on linkages and public understanding?
## (Preliminary White Paper, February 11, 2022)

Reza Abdi and Terri S. Hogue[1]

## Abstract

Wildfires are increasing in frequency and size across the western U.S., with some of the deadliest fires in recorded history occurring in the last few years. The public, as well as elected officials, use social media to convey opinions and knowledge on topics that are impacting their communities. We utilize the platform Twitter to assess connections of wildfire to climate change during recent events and evaluate the differences in knowledge between the public and their government officials. Results show some linkages of wildfire cause and effect, although this relationship was not large (only 5%) and was even lower at the governmental level (2%), suggesting that a broader number of the public and government did not relate climate change to recent extreme wildfires.

**Index terms:** Wildfires, Climate change, Natural Language Processing, Social Media Mining, Twitter

1. Introduction

The frequency of wildfires in the western US has increased by 400% since 1970 [1]. In 2018, the Camp Fire in Northern California devastated the town of Paradise, killing 85 people, making it the deadliest fire in California history. The Camp Fire burned 153,336 acres and destroyed 18,733 buildings. It was also the world's costliest natural disaster in 2018, causing $16.5 billion in damage [2]. To date (as of 9/15/2021) large fires have burned more than 5.6 million acres in the US. During the same time period (1/1/20-9/15/20), in 2020, more than 6.8 million acres have burned in the US, 8% more than the 10-year average and almost 60% more than the area burned in 2019 [3]. Reported wildfires consumed 10.1 million acres nationally in 2020, 2.2 times more compared to 4.6 million acres in 2019 [3-1]. A large proportion of the burned acreage occurred in California, accounting for 38% of the nation's total burned acres. In 2020, the reported number of wildfires for the US was well above 10-year averages in all Geographic Areas, except in Alaska (65%), Southern Area (60%), and Rocky Mountain Area (95%). The remaining Geographic Areas in the US were well above average: Eastern Area (138%), Great Basin (117%), Northern California (122%), Northern Rockies (121%), Northwest (115%), Southern California (123%), and Southwest (121%) [3-1].

Almost 670,000 acres of the state of Colorado's forest have been burned in 2020. Year 2020 was the most active fire season ever recorded in Colorado and includes the three largest fires in the state's history [3-2]. Wildfires in the state of Oregon burned approximately 1 million acres of land, almost double the 10-year average of 557,000 acres [4]. These current trends are expected to continue and also continue to impact the wildland-urban interface (WUI) [5] as for example, during Oct. 2020, 300-400 structures were lost in the East Troublesome Fire in and around Grand Lake, Colorado [6] and during 2020 more than 2200 residences were destroyed in Oregon [6].

In scientific communities, the linkages between wildfire, climate change, and forest management are generally known [7, 8]. A changing climate is driving a longer forest fire season with larger and more severe fires [9]. Climate change causes rising air temperature and increased evaporation, leading to drier soil and more flammable vegetation [10]. Further, research has confirmed that in a changing climate winter snowpack melts about a month earlier, forests are now enduring longer dry seasons after snowmelt passes [11, 12]. The fire season could shift from fall into winter, with longer and more intense fires [13]. Human-made climate effects on wildfire can vary greatly across space and time due to confounding factors such as natural climate variations, land and fire management practices, ignitions from humans, spatial diversity in vegetation type, and the complex ways in which these processes interact [14, 15].

---

[1] *Reza Abdi and Terri Hogue are with the Department of Civil and Environmental Engineering, Colorado School of Mines, Golden, CO, USA 80401; Corresponding author's email (i.e., Reza Abdi): {rabdi}@mines.edu*

Politicians and the public can also express strong opinions on wildfire, climate change, and forest management linkages, but the degree of this connection is unknown. Social media and machine learning-based (ML) techniques offer resources and tools to quickly assess public and government officials' responses to wildfires and their potential connection to climate change. Hence, we address the following questions in this research: 1) How often were recent wildfires connected to climate change on social media platforms? and 2) Are there differences between public and government officials on climate change and wildfire connections? We advocate that results provide insight on societal understanding of climate-wildfire connections and also provide a pathway for accessing potential areas for education and outreach efforts on environmental issues.

2. Methods

Twitter was utilized to analyze public and elected government official statements on wildfires and climate change, including use of sentiment analysis and topic modeling. The Twitter developer application programming interface (API) in Python's Tweepy package [16] was used to stream tweets within the US from 9/3/2020-9/23/2020 when wildfires were dominating the news headlines in the US. Keywords utilized included wildfire and forest fire and a total of 185328 tweets were extracted (retweets were not included). We also obtained all tweets, as well as their likes and retweet numbers, from three governmental-type accounts for the same period from three states that experienced wildfires during this period: California, Colorado, and Oregon. We extracted local (cities and/or counties dealing with wildfires, 15 accounts), state-level (three governors and 15 representatives in the determined districts, 18 accounts), and federal level (senators, 6 accounts) tweets for the three states, for a total of 39 accounts and 4086 tweets (see the supplementary material for more details); 488 of these tweets were related to wildfire (See Fig. 1).

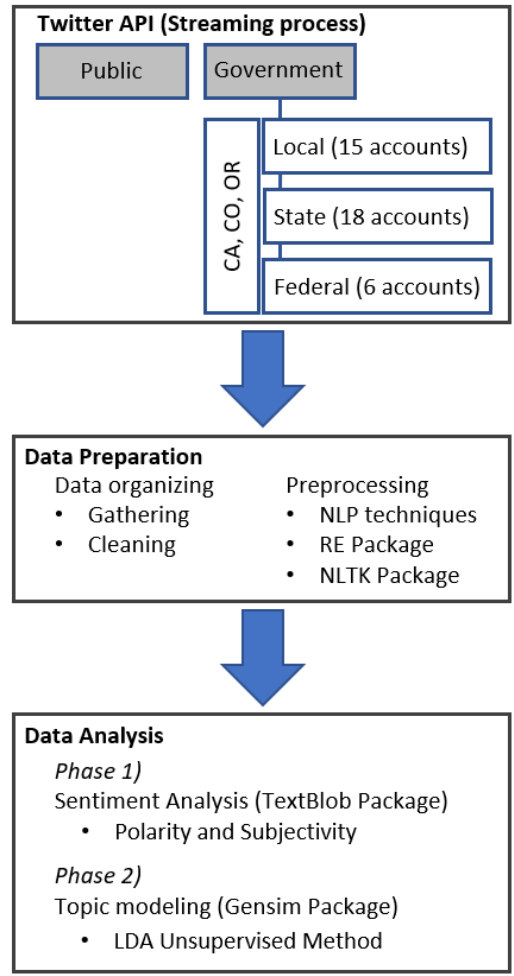

Figure 1. Conceptual diagram explaining the process we applied for mining the Twitter social media accounts to stream, process, and analyze the data (tweets).

The contents of the streamed tweets were preprocessed for use by the Python libraries and packages, for text mining, natural language processing (NLP) techniques, and algorithms. We utilized Python libraries, regular expression (RE) [17], and natural language toolkit (NLTK) [18] functions for preprocessing (see Table S1 for more details). Sentiment analysis was undertaken to assess the polarity and subjectivity of the tweets. We used the TextBlob [19] and NLTK libraries and applied the tokenization process. The resulting polarity score is a float number that ranges from -1 to +1, where -1 is a negative statement and +1 means a positive statement. Subjective tweets refer to personal opinion, emotion, or judgment whereas objective tweets refer to factual information. Subjectivity is also a float number that ranges from 0 and +1; +1 implies public opinion and not factual information. We applied a Latent Dirichlet Allocation (LDA) [20] unsupervised methodology on tweets for topic modeling, using Gensim [21], Scikit-Learn [22], NLTK, and spaCy [23] libraries. Topic modeling was also utilized to discover abstract topics that occur in the constructed dataset. This is an unsupervised machine learning technique in NLP that does not require training and is frequently used for the discovery of hidden semantic structures in a text body.

3. Results and discussion

Wildfire events were reflected in social media reactions among both public and governmental accounts in the collected data. The number of tweets with the tag "wildfire" in the public accounts rose sharply - from 1627 tweets on Sep. 5th to 15988 tweets on Sep. 11th. The average number of tweets with the tag wildfire was 1740 in the prior Sep. 3rd and Sep. 4th window (Fig. 2a). There was a relatively similar pattern observed between the public and governmental tweets related to wildfire (Fig. S1a[2]), where the number of tweets increased from 1 tweet on Sep. 5th to 51 tweets on Sep. 11th for the governmental accounts where the average number of tweets prior to the extensive wildfires was 5 tweets between Sep. 3rd and Sep. 4th. We observed a relatively similar pattern in tweets with the tag word climate change with wildfire (Fig. 2b and Fig. S1b).

We also saw a relatively similar increase in tweets with the use of "climate change", from 19 to 680 tweets from Sep. 5th to Sep. 11th (Fig. 2b and Fig. S1b). The number of tweets about climate change reached its maximum among both public and governmental accounts the day after President Trump's California visit on Sep. 14th (Fig. 2b). The number of tweets with climate change was 9244 out of 185328 tweets (5%) in the public dataset and 65 tweets out of 4086 tweets (2%) in the governmental dataset (See Fig. S2, Fig. S3, and Fig. S4 for more details).

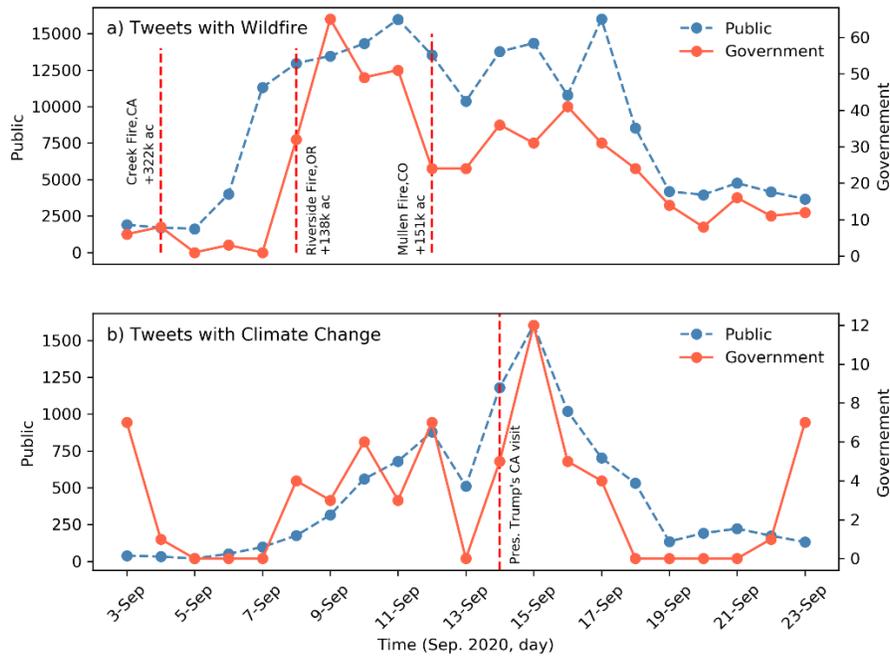

Figure 2. Variation in the number of tweets containing the tag word wildfire (a), and climate change within the same wildfire tweets (b).

---

[2] *Figures and Tables with the "S" refers to the materials that are presented in the supplementary materials file.*

Extracting tweets containing climate change from the government preprocessed dataset (Fig. S5) showed that in 70% of the comparisons the number of likes and retweets mostly were significantly higher ($p < 0.05$) compared to the tweets without the climate change tag word (see Table S2 for more details). The median of the total number of tweets for local, state, and federal levels that included climate change text received 16 and 4.8 times more likes and retweets respectively, compared to tweets about other topics, indicating that Twitter users paid more attention to tweets concerning climate change than other topics being tweeted (i.e. COVID-19, BLM, etc.; See Fig. 3).

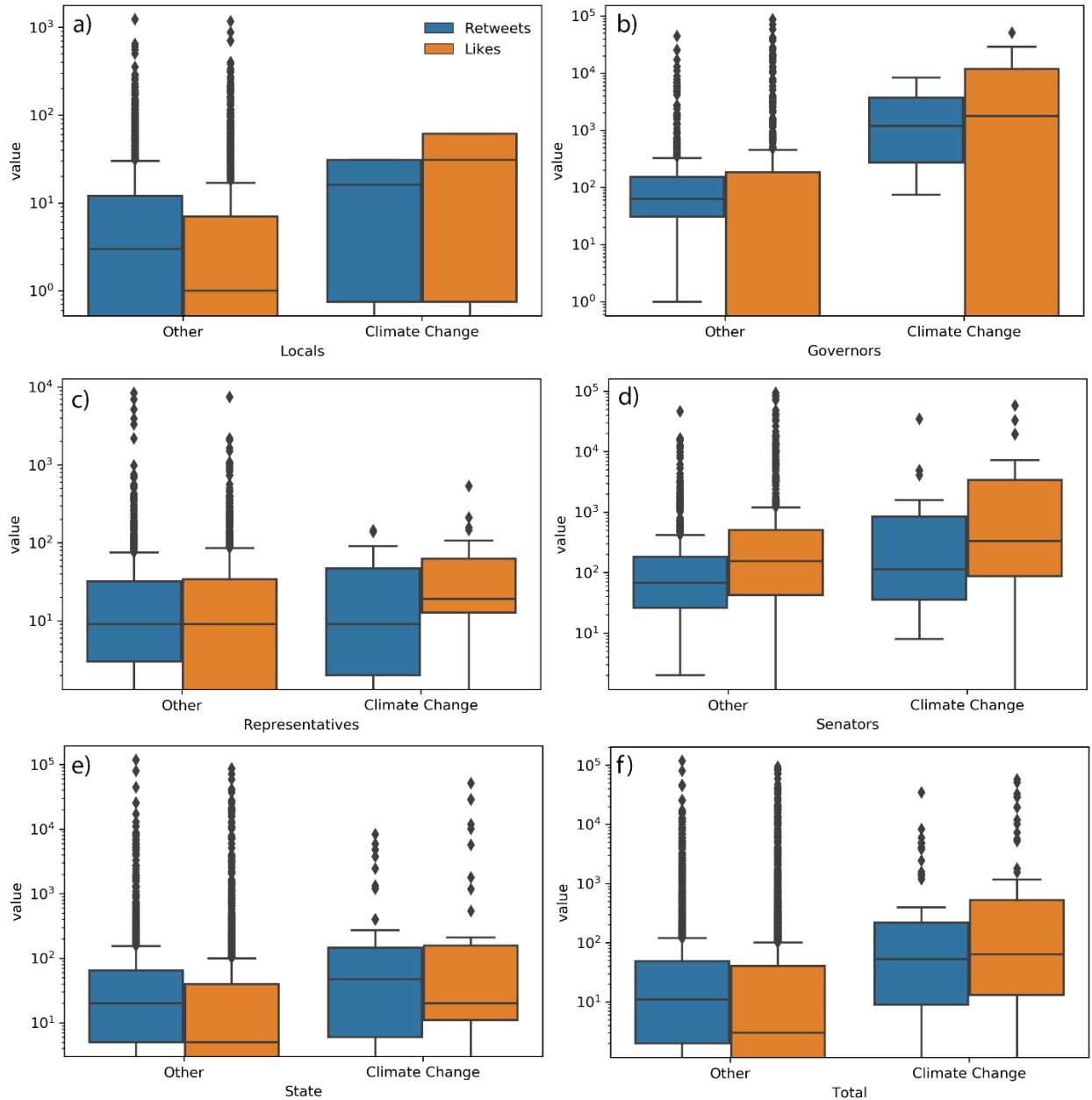

Figure 3. Boxplots showing the number of likes and retweets for the tweets with tag word "climate change" against tweets with other topics for the local level (a), state level, only the governors (b), state level, only the representatives (c), federal level, the senators (d), state-level with both governors and representatives (d), and total tweets including all the local, state, and federal levels (f).

Sentiment analysis of the public dataset demonstrated that tweets related to climate change displayed more negative feelings. The polarity of the tweets with climate change-related to wildfire had a 25[th] and 75[th] percentiles of -0.1 and 0.02, respectively. In contrast, these numbers were 0 and 0.1 respectively for all wildfire tweets during the same time period (Fig. 4 and Fig. S6). We hypothesize that the slightly positive feelings in wildfire-related tweets were likely due to the extensive efforts related to fighting the wildfires, especially attributed to the US Forest Service and the California Department of Forestry and Fire Protection department. The subjectivity of the tweets with climate change was more objective with a median of 0.1, compared to 0.25 in the wildfire dataset.

Using the unsupervised LDA topic modeling method on tweets with climate change, we extracted four topics. A total of 38% of the topics were related to wildfires and climate change; within this subject, people were also discussing forest management (Fig. S7). The second most important topic (25% of the tokens) was related to the risks of climate change as a potential crisis. The interactive HTML format of pyLDAvis visualization is downloadable through the following link:
https://github.com/reabdi/WildFiresTopicModeling.git

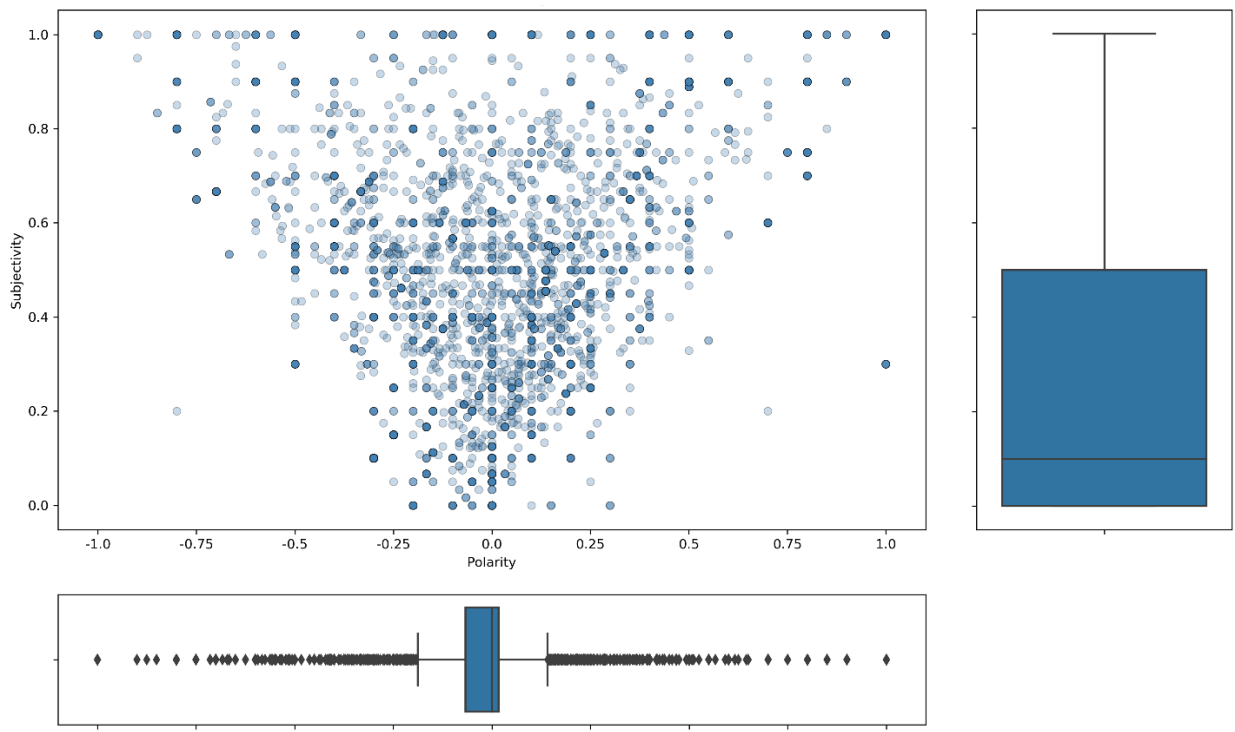

Figure 4. Distribution of sentiment subjectivity over sentiment polarity of the public tweets dataset for wildfire tweets that include climate change (9244 tweets). Sidebars show the individual distribution of the sentiment subjectivity and sentiment polarity for the tweets.

The topic modeling analysis illustrated that in around 60% of tweets with climate change, people made the connection between wildfires being driven by climate change. Similarly, we saw the same pattern in public tweets after extreme wildfire events (Fig. 1a). On the day after President Trump's California visit on Sep. 14[th], both the public and governmental data showed the maximum number of tweets related to climate change and wildfires (Fig. 1b); again, demonstrating that both groups reacted to the President's statements about how climate change was not causing wildfires.

Even though we saw some linking of wildfire cause and effect, the number of tweets pointing to this relationship was not large (only 5%). It was even lower at the governmental level (2%), suggesting that a broader number of the public and government did not mention climate change as the cause of the extreme wildfires. On the other hand, the higher number of likes and retweets of the tweets at the governmental level referring to climate change as a scientific cause for the wildfires demonstrates that at least in the three studied states, the public reacted more actively to the tweets referring to wildfires as consequences of climate change. Putting these facts about the behavior of the public and governmental levels together suggests the importance of increasing environmental awareness among both the public and their political representatives. Further, the results will inform opportunities for education/outreach-focused activities through social media platforms like Twitter for governmental officials to transfer their messages to a broad audience.

Social media platforms such as Twitter, which has shown significant use among both public and governmental sectors in recent years, could be used as a powerful platform by relevant groups to educate the public and transfer relevant messages to a wide range of users. These efforts could be further assessed to see if the transferred knowledge has a tangible effect on people's opinions on climate change. However, initially, the government institutes at local, state, and federal levels should be more engaged and informed about the role of climate change on wildfires and potentially other catastrophes like droughts, heatwaves, and sea-level rise. The latter could be achieved with efforts from the scientific societies, providing strong evidence and factual data and findings.

4. Conclusions

In this study, we utilized NLP and text mining tools and knowledge, typically used by data or computer scientists, for a socio-environmental objective in order to analyze the public and government opinions using the social media platform, Twitter, to investigate relationships between wildfires and climate change. Based on this multidisciplinary research plan and using Twitter API and Python's Tweepy package, we streamed tweets with tag words wildfire and forest fire from Sep. 3rd to Sep. 23rd, 2020. We selected this time window based on the number of wildfire events at the time across the western US. We streamed the public tweets from across the US and government-related tweets in three levels of local, state, and federal accounts (39 accounts in total) in states California, Colorado, and Oregon, three states that were experiencing spikes in the number and domination of the wildfire events in the 2020 fire season.

Our analysis showed that both public and government didn't have broad knowledge regarding the linkages between wildfires and climate change. The results showed only 5% and 2% in the public and government tweet datasets respectively were referring the climate change as a potential reason for the wildfires. However, we found that in that limited number of the tweets from the government accounts that were related to the connections between wildfires and climate change, the number of the likes and retweets were in general significantly higher compared to other topics. This attention from the public to the government tweets pointing to wildfires as a consequence of climate change reveals that the linkages between wildfires and climate change should be transferred via educating the public. Wildfire, climate change, and other related concepts can be emphasized in social media as inexpensive and effective platforms to improve environmental awareness among the people. Due to their roles in society, government accounts could be good resources for the public to transfer these messages.

**Acknowledgments**

The authors would like to warmly thank Dr. Ashley Rust for reading this work and providing us with productive comments.

**Supplementary Materials**

Table S1. The list of the 30 most frequent tag words extracted from the constructed text corpus after the preprocessing and tokenization process.

| # of words after preprocessing and tokenization process | First 30 most frequent tag words in the text corpus |
|---:|---:|
| 95640 | wildfire |
| 26456 | smoke |
| 25402 | forest |
| 21430 | California |
| 12551 | Oregon |
| 12508 | like |
| 10725 | gender |
| 9788 | reveal |
| 9366 | fires |
| 9301 | wildfires |
| 9244 | climate |
| 7301 | people |
| 7106 | west |
| 6348 | air |
| 5437 | party |
| 5396 | amp |
| 5258 | Trump |
| 5079 | change |
| 5050 | coast |
| 4666 | state |
| 4063 | started |
| 4019 | season |
| 4014 | time |
| 4009 | covid |
| 3821 | today |
| 3805 | new |
| 3746 | national |
| 3598 | spread |
| 3578 | know |
| 3477 | news |

Table S2. The 10[th] percentile, median, and 90[th] percentile of the number of likes (favorite) and retweets of the governmental accounts separated to local (cities and counties), state (representatives and governors), and federal level (senators) for the tweets with climate change against all the other topic.

| Category | Statistic Value | Other | | Climate Change | |
|---|---|---|---|---|---|
| | | Retweets | Likes | Retweets | Likes |
| Locals (Cities and Counties) | 10[th] percentile | 0 | 0 | 0 | 0 |
| | Median | 3 | 1 | 15 | 31 |
| | 90[th] percentile | 36 | 31 | 31 | 61 |
| State (Representatives) | 10[th] percentile | 1 | 0 | 1 | 0 |
| | Median | 10 | 9 | 9 | 19 |
| | 90[th] percentile | 95 | 108 | 72 | 149 |
| State (Governors) | 10[th] percentile | 13 | 0 | 99.6 | 0 |
| | Median | 63 | 0 | 1192 | 1788 |
| | 90[th] percentile | 597 | 2063 | 5708 | 28944 |
| State (Both) | 10[th] percentile | 2 | 0 | 2 | 0 |
| | Median | 20 | 5 | 47 | 20 |
| | 90[th] percentile | 176 | 231 | 2449 | 10086 |
| Federal (Senators) | 10[th] percentile | 9 | 0 | 13 | 22 |
| | Median | 68 | 154 | 113 | 331 |
| | 90[th] percentile | 1032 | 3270 | 4695 | 19495 |
| Total | 10[th] percentile | 0 | 0 | 2 | 0 |
| | Median | 11 | 4 | 53 | 64 |
| | 90[th] percentile | 155 | 273 | 3888 | 14802 |

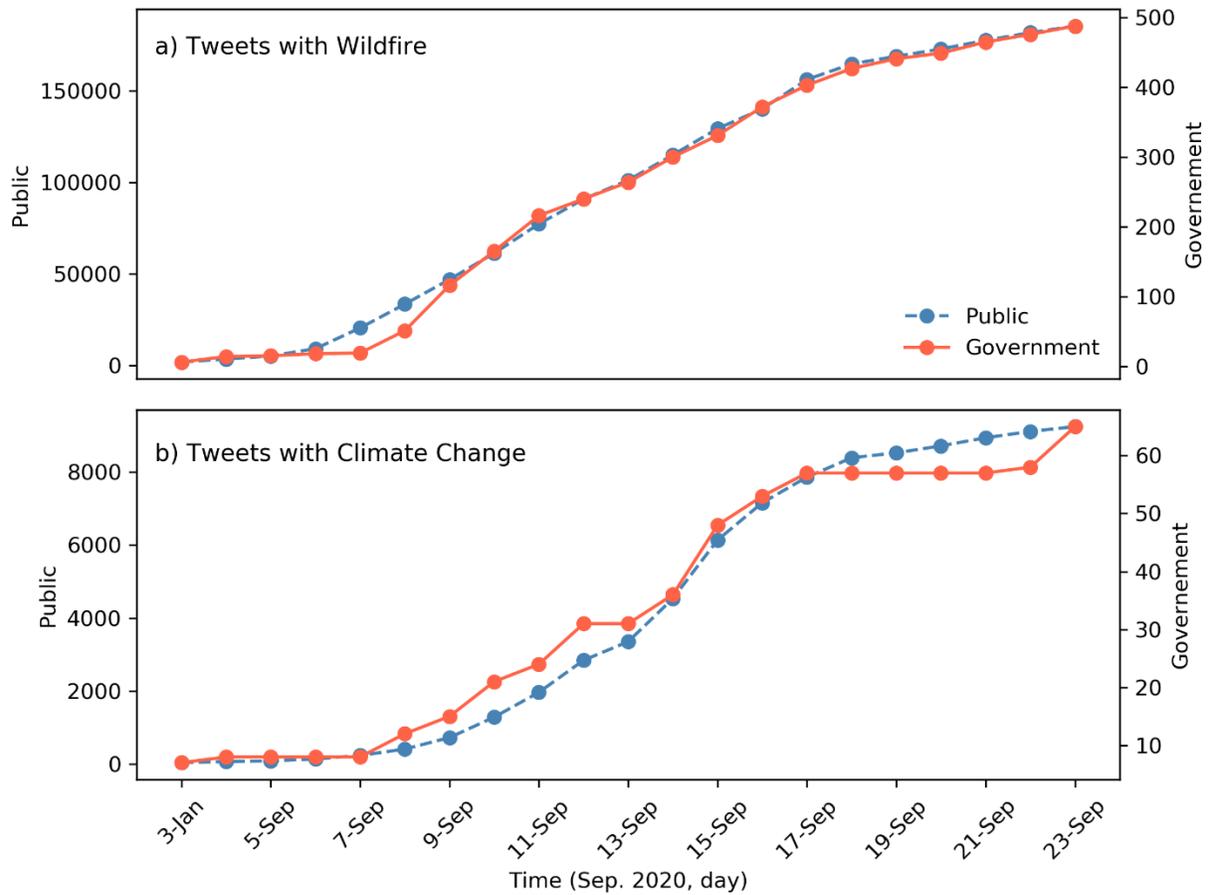

Figure S1. Accumulated number of the tweets containing the tag word "wildfire" (a), and "climate change" within the tweets with tag word "wildfire" (b)

Figure S2. Word cloud made form public tweets that used "wildfire" in 185328 tweets

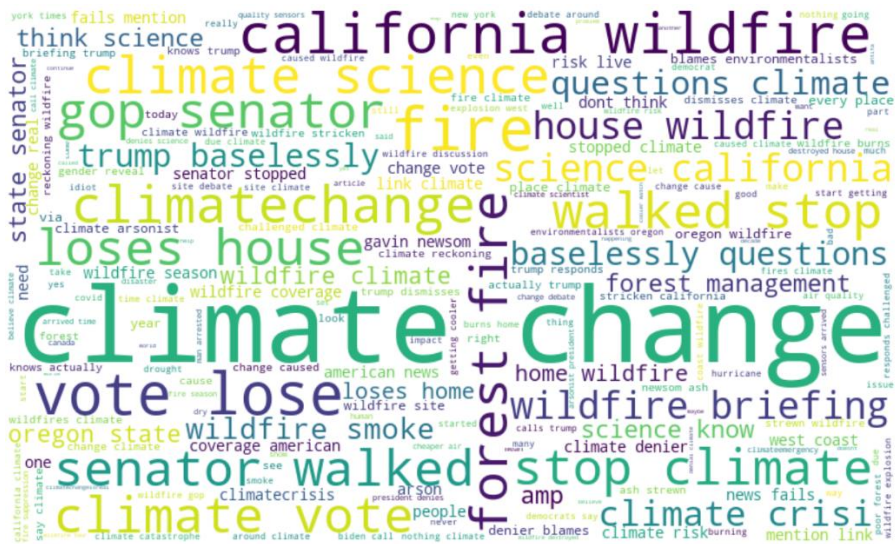

Figure S3. Word cloud made form public tweets that used "climate change" in 9244 tweets

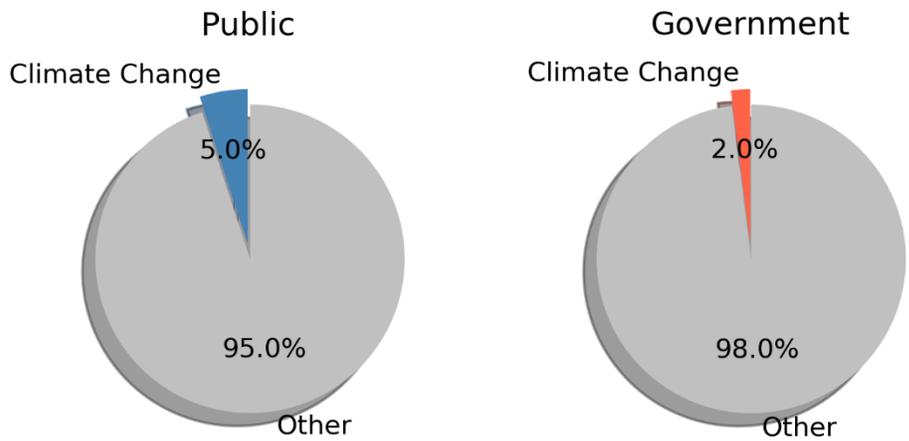

Figure S4. Percentage of the tweets with the tag word "climate change" in the public and governmental tweet datasets.

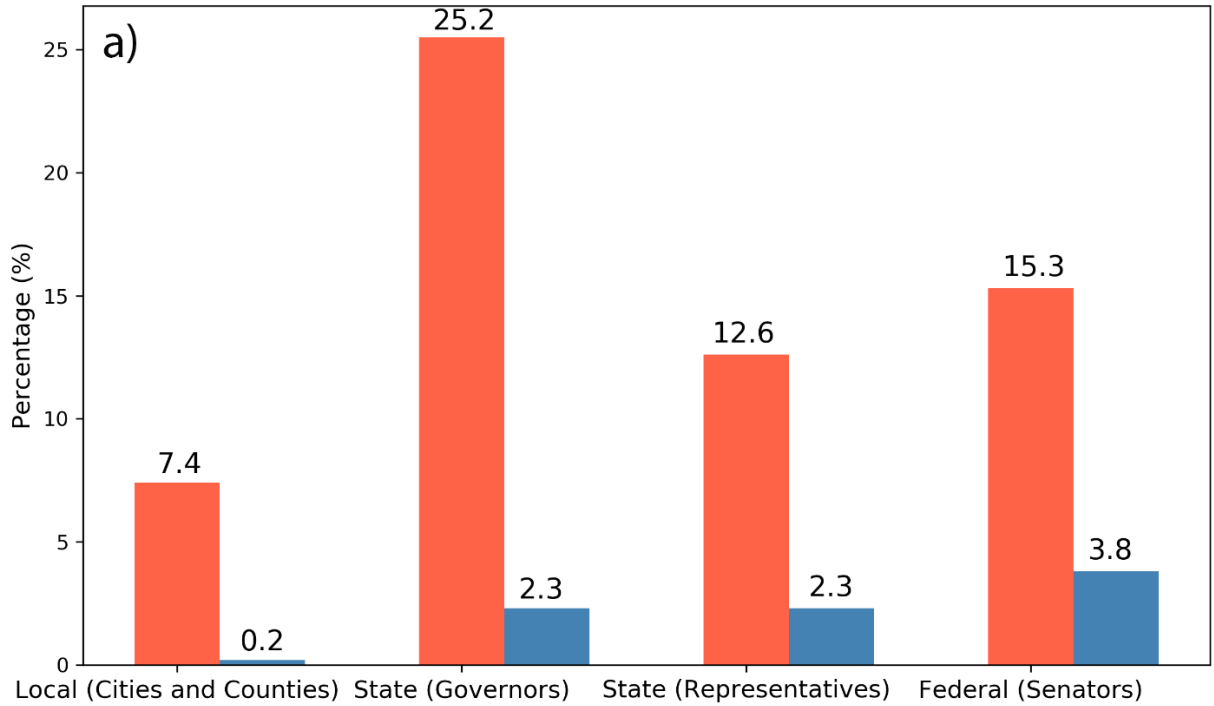
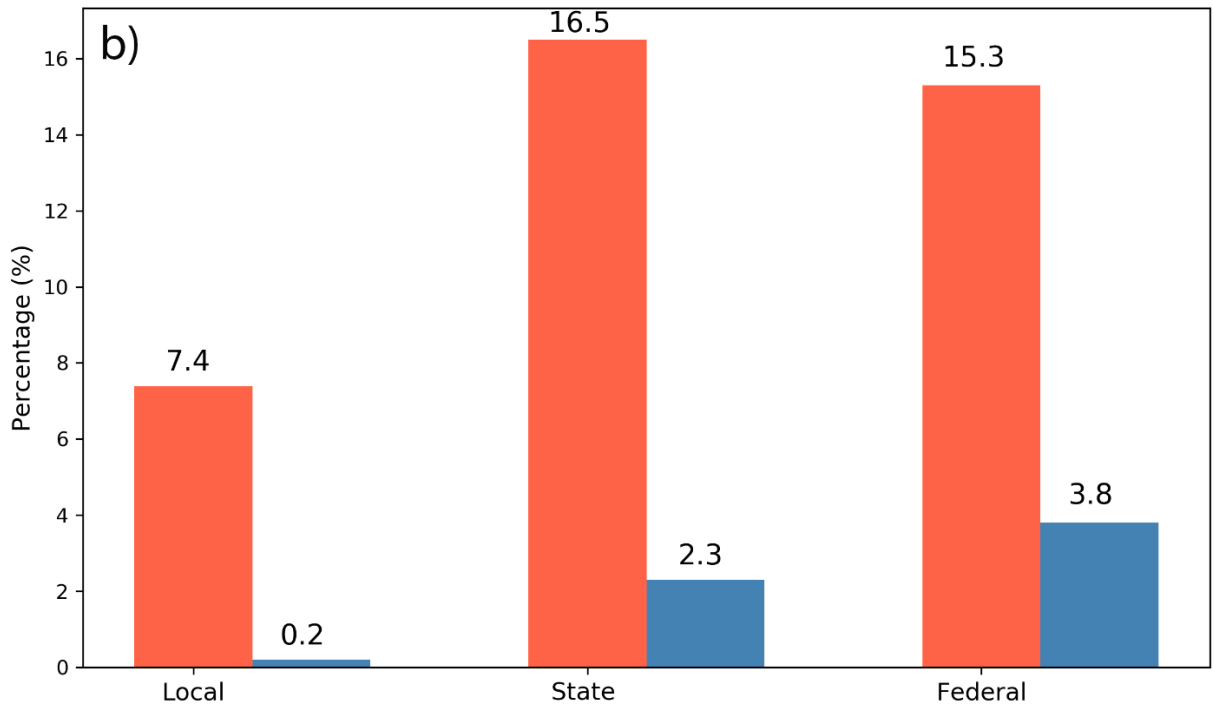

Figure S5. The percentage of using tag words "wildfire" and "climate change" in the governmental tweet dataset separated based on their level. Panel (a) shows the state-level separately for the governors and representatives and panel (b) shows those two categories combined.

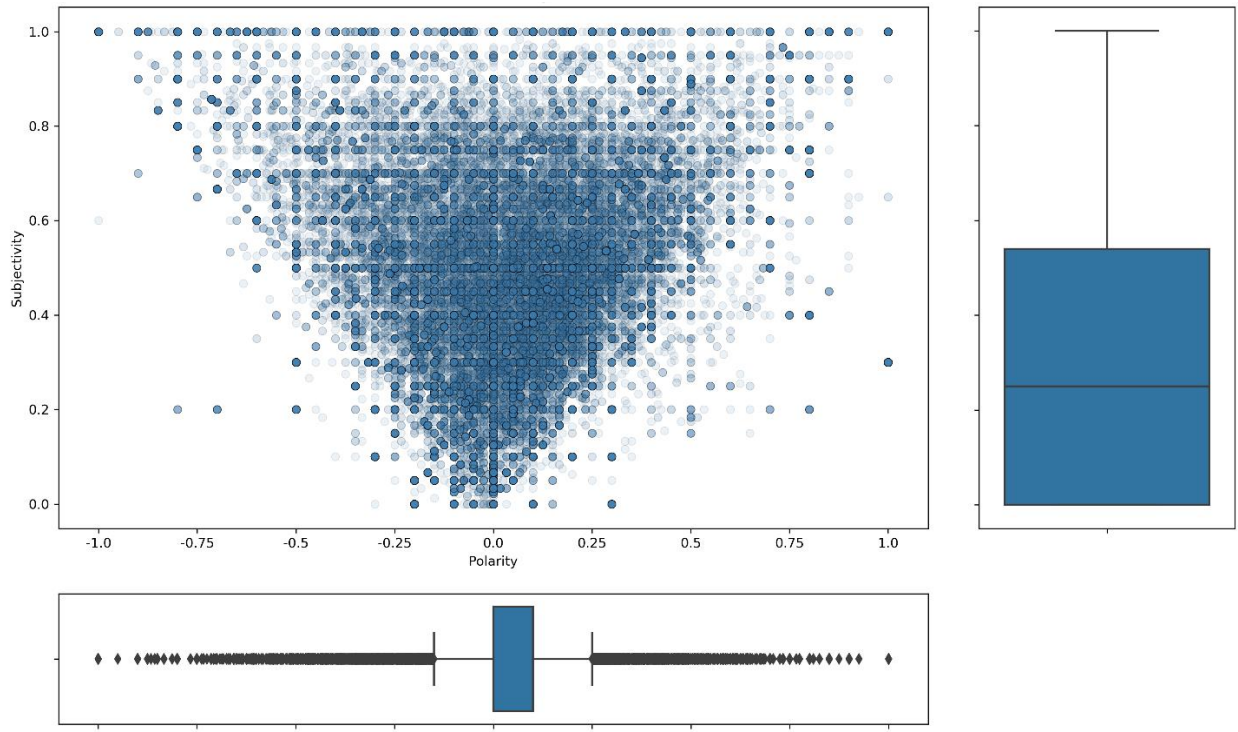

Figure S6. Distribution of sentiment subjectivity over sentiment polarity of the public tweets text corpus for the tweets with the tag word "wildfire" (185328 tweets). The sidebars show the individual distribution of the sentiment subjectivity and sentiment polarity.

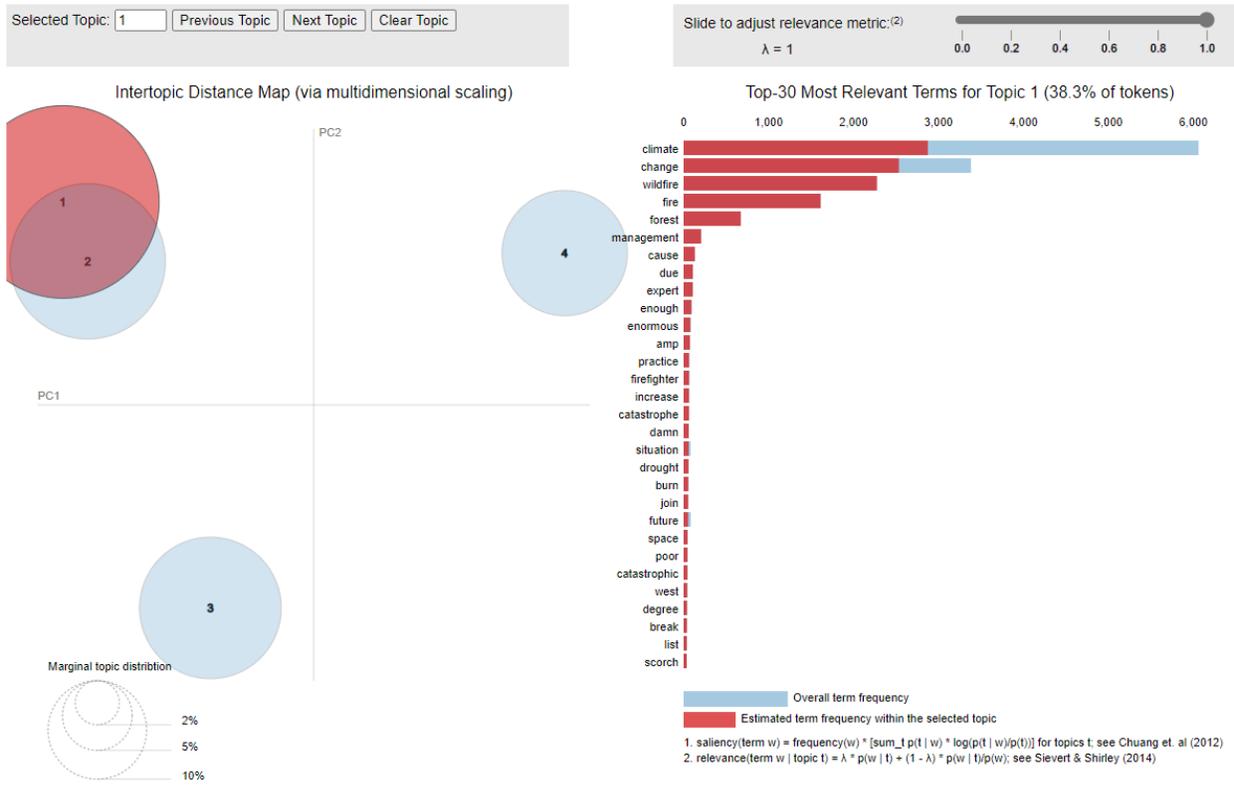

Figure S7. Screenshot of the unsupervised LDA topic modeling visualization, showing the most important topic identified from the tweets with climate change tag word by the LDA algorithm.

Detailed list of the Twitter accounts we have used as the governmental level for our analysis:

Locals:
CA: Counties of Napa, Mendocino, Santa Clara, Sonoma, and Fresno
OR: City of Salem, and Counties of Lane, Clackamas, Jackson, and Multnomah
CO: County of Larimer and Boulder Cities of Boulder, Grand Junction, Glenwood Springs, and Cortex
15 accounts in total

State-level:
Governors of three states, California, Colorado, and Oregon
Congress representatives: 5 for each state
18 accounts in total

Federal: 6 senators, California, Colorado, and Oregon